\documentclass[twocolumn,preprintnumbers,amsmath,amssymb,prl]{revtex4}

\usepackage{graphicx}
\usepackage{amssymb}

\begin{document} 

\title{Hyper-gratings: nanophotonics in planar anisotropic metamaterials}

\author{Sukosin Thongrattanasiri}
\author{Viktor A. Podolskiy}
\email{viktor.podolskiy@physics.oregonstate.edu}
\affiliation{Physics Department, Oregon State University, 301 Weniger Hall, Corvallis, OR 97331, USA}

\date{\today}

\begin{abstract}
We present a technique capable of producing subwavelength focal spots in the far-field of the source in planar non-resonant structures. The approach combines the diffraction gratings that generate the high-wavevector-number modes and planar slabs of homogeneous anisotropic metamaterials that propagate these waves and combine them at the subwavelength focal spots. In a sense, the technique combines the benefits of Fresnel lens, near-field zone plates, hyperlens, and superlens, and at the same time resolves their fundamental limitations. Several realizations of the proposed technique for visible, near-IR, and mid-IR frequencies are proposed, and their performance is analyzed theoretically and numerically. Generalization of the developed approach for sub-diffractional on-chip photonics is suggested. 
\end{abstract}


\maketitle 

The further development of sensing, imaging, and communication technologies requires ever improving control over the propagation of electromagnetic waves\cite{shalaevNature,pendryTransform,superlens,hyperlens,stockmanPRL,nfZonePlates,zhangLayers}. A generic photonic unit (e.g. optical sensor, lithographic or imaging apparatus, communication unit) can be considered as a device providing the optical communication between several spatially separated spots. The separation between the spots (``focal distance'') and the size of the spot (resolution) are among the main figures of merit for optical systems\cite{superlens}. Here we propose the technique for {\it far-field communications} between several {\it sub-wavelength} spots based on manipulation of sub-wavelength signals in {\it planar slabs} of strongly anisotropic dielectrics. Our method combines the benefits of planar optics offered by Fresnel zone plates\cite{bornWolf} and negative-refraction lenses\cite{veselago,superlens}, wide-spectrum-generation offered by near-field plates\cite{nfZonePlates}, and diffraction-less propagation offered by strongly anisotropic (``hyperbolic'') metamaterials\cite{podolskiyPRB,hyperlens,indefinite}. The method is illustrated on the example of far-field sub-wavelength foci generated by the diffraction gratings in hyperbolic metamaterials. Analytical estimates of the performance of such ``hyper''-gratings are provided and verified with numerical solutions of Maxwell equations. Generalizations of the proposed technique for on-chip communications are suggested.

\begin{figure}[thb]
\centerline{\includegraphics[width=8cm]{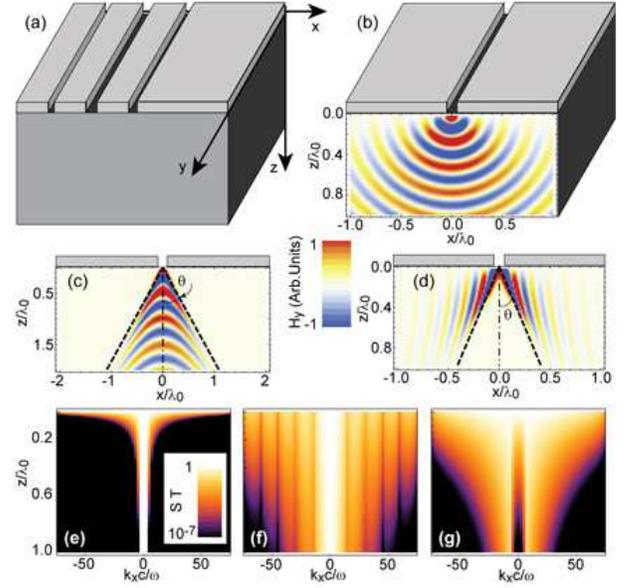}}
\caption{
(a) (color online) schematic of the hyper-grating device; $1D$ grating is shown; (b) propagation of the diffraction-limited pulse in isotropic material; $\lambda_0=15\mu m; d_1=200 nm; x_1=0; \epsilon_z=\epsilon_{xy}=16$\cite{footnoteSiTheBest}; 
(c,d) sub-diffraction propagation in nanowire(c) and nanolayer(d) anisotropic metamaterials; $d_1=100 nm$ (c) and $d_1=200 nm$ (d); dashed lines show the direction of $|k_x c/\omega|\gg 1$ modes [Eq.\eqref{eqTheta}]; panels (e{\ldots}g) show field spectra of the systems in (b{\ldots}d) respectively
} 
\end{figure}

The optical behavior of imaging or focusing devices is most clearly seen when considered in the wavevector space. In this approach, the optical pulse at the entrance to the device is represented as a set of plane waves with well-defined components of the wavevector $\vec{k}$ and frequency $\omega$. The spatial evolution of each of these waves is then analyzed, and finally the spectrum is converted back into the real-space domain at the exit point of the device. 

Most transparent natural materials support propagating waves with some limited range of the wavevectors. The maximum value of the wavevector component 
$
k_x^{\rm max}=\sqrt{\epsilon\mu} {\omega}/{c}
$
determines the minimum size of the focal spot that can be achieved in this device in the far-field limit $\Delta\sim 2\pi/k_x^{\rm max}=\lambda_0/\sqrt{\epsilon\mu}$\cite{bornWolf}. The role of the conventional optical elements is therefore reduced to adjusting the phase-shifts between different wavevector components to achieve the best-possible (although still diffraction-limited) resolution. This adjustment is typically provided by material inhomogeneities (lenses), or by diffraction gratings (Fresnel lenses or zone plates)\cite{bornWolf}. 

Several techniques have been suggested to achieve subwavelength imaging. Some of these techniques -- SNOM, superlens, and near-field plates\cite{natureSNOM,superlens,nfZonePlates} rely on exponentially decaying (evanescent) fields with $|k_x|>k_x^{\rm max}$ to surpass the diffraction limit. Unfortunately, the realistic applications of these techniques are limited to near-field proximity of the imaging system\cite{podolskiyOL}. 

Another class of structures, either use transformation-optics techniques (light compressors)\cite{pendryTransform}, or make use of ultra-high-index modes in plasmonic\cite{stockmanPRL} or strongly anisotropic\cite{podolskiyPRB} media. While these systems are able to achieve subwavelength light manipulation in the far-field, their fabrication requires three-dimensional patterning. Moreover, the devices themselves are often non-planar which further restricts the range of their possible applications.

\begin{figure}[t]
\centerline{\includegraphics[width=8cm]{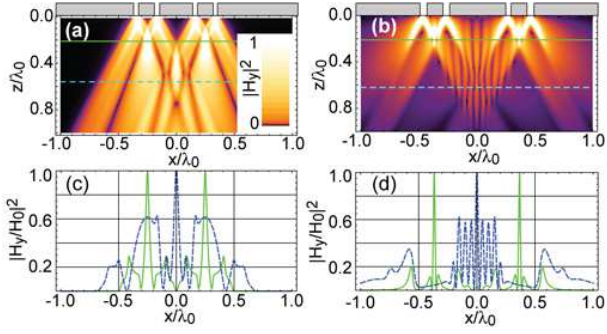}}
\caption{
(color online) $1D$ hyper-gratings for sub-diffractional photonics in the bulk of nanowire- (a) and nanolayer-metamaterials (b); profile of several sub-wavelength foci from (a) and (b) is shown in (c) and (d) respectively; the fields in (c,d) are normalized to the maximum field at the given distance from the grating; dashed and solid lines in (a,b) correspond the positions of field profiles shown in (c,d);  
geometry parameters: (a,c): $d_1=d_2=d_3=d_4=100 nm$, $x_1=-500nm, x_2=-300nm, x_3=300 nm, x_4=500 nm$; (b,d): $d_1=d_2=d_3=d_4=200nm$, $x_1=-7\mu m, x_2=-4\mu m, x_3=4\mu m, x_4=7\mu m$
} 
\end{figure}

Here we propose a system capable of far-field sub-wavelength light manipulation free of the above restrictions. The schematics of the proposed planar structure and several realizations are shown in Fig.1a. The system comprises a (planar) slab of strongly anisotropic metamaterial covered with a diffraction grating. As explained below, the diffraction grating is responsible for generating the high-wavevector components of electromagnetic fields, and the slab is used for the routing of the resulting sub-wavelength signals. 

We begin by discussing the propagation of TM-polarized light generated by a thin slit of width $d_1\ll\lambda_0=2\pi c/\omega$ positioned at $x=x_1$. We assume that the pulse propagates along the optical axis (axis $z$ in Fig.1a) of a uniaxial anisotropic metamaterial with dielectric permittivity $\hat{\epsilon}$, a diagonal tensor with components $\{\epsilon_{xy},\epsilon_{xy},\epsilon_z\}$. The field due to the slit inside the material at the distance $z$ from the interface is given by: 
\begin{equation}
\label{eqFld}
H_y(x,z)=\int_{-\infty}^{\infty}S(k_x)T(k_x,z)e^{i k_x x}dk_x,
\end{equation}
with the source [$S(k_x)$] and the transfer [$T(k_x,z)$] functions given by: 
\begin{eqnarray}
\label{eqSfun1}
S(k_x;x_1,d_1)&=&2\frac{\sin(d_1\; k_x/2)}{k_x}e^{-ik_x x_1}
\\
\label{eqTfun1}
T(k_x,z)&=&\exp\left[\pm i \sqrt{\epsilon_{xy}\left(\frac{\omega^2}{c^2}-\frac{k_x^2}{\epsilon_z}\right)} z\right]
\end{eqnarray}
The sign of the square root in Eq.\eqref{eqTfun1} should be chosen to enforce the field decay inside the absorptive medium\cite{veselago}. 

The amplitude of the transfer function defines the evolution of the wavevector spectrum, and thus it effectively defines the resolution of the system. As seen from Eq.\eqref{eqTfun1}, in isotropic systems, the high-$k_x$ components of the spectrum are exponentially suppressed [Fig.1(b,e)], and the sub-wavelength resolution is limited to the proximity of the slit. 

The situation is dramatically different in strongly anisotropic metamaterials that have $\epsilon_{xy}\epsilon_z<0$ where hyperbolic dispersion virtually eliminates high-$k_x$ cut-off\cite{podolskiyPRB}. Nanolayer\cite{hoffman} and nanowire\cite{podolskiyWires,zayatsWires,zhangWires} realizations of these unique structures have been theoretically predicted and experimentally demonstrated for near-UV, visible, near- and mid-IR frequencies. It has been shown -- both theoretically and experimentally -- that optical properties of relatively thick metamaterials are well-described by (local) effective medium theories\cite{podolskiyWires,hoffman,zhangWires}. In this work we illustrate the sub-diffractional manipulation of light on examples of Au-alumina nanowire system\cite{zayatsWires,zhangWires,podolskiyWires} and AlGaAs-InGaAs nanolayer\cite{hoffman} structures operating at $1.5\mu m$ and $15 \mu m$ respectively, where applicability of effective medium theories has been recently verified. The particular material parameters that we use in our simulations are: $\epsilon_{xy}=3.6+0.005i; \epsilon_z=-12.2+1.36i$ for nanowire system and $\epsilon_{xy}=-6.4+1.4i; \epsilon_z=36+3.4i$ for multilayered structure\cite{podolskiyWires,palik}. 

Note that the two structures have opposite anisotropy. The nanowire composite exhibits the negative refraction (positive phase index) properties, described in\cite{hoffman,zhangWires,indefinite,veselago}, and supports the propagation of both small ($|k_x c/\omega|\lesssim 1$) and high-wavenumber waves. In contrast to this behavior, the nanolayer composite operates in the positive-refraction (negative index) regime; it only supports high-wavenumber components\cite{podolskiyPRB,veselago}. As seen in Fig.1, both structures dramatically outperform their isotropic counterparts. However, suppression of diffraction-limited background in nanolayered structure makes these systems more suitable for far-field operations. 

As described above, the sub-diffraction manipulation of light relies on the propagation of high-$k_x$ modes. In strongly anisotropic systems these waves propagate almost parallel to each other, with the angle between propagation direction and optical axis given by the ratio of wavevector components\cite{prime}: 
\begin{equation}
\label{eqTheta}
\tan\theta=\frac{ \sqrt{\epsilon_{xy}\left(\frac{\omega^2}{c^2}-\frac{k_x^2}{\epsilon_z}\right)}}{ k_x }
\simeq\sqrt{-\frac{\epsilon_{xy}^\prime}{\epsilon_{z}^\prime}}.
\end{equation}
The concentration of sub-wavelength components into the two sub-wavelength beams emerging from the point slits are clearly seen in Fig.1; the directions of these beams are in perfect agreement with Eq.\ref{eqTheta}. Note that while the field distributions in nanowire metamaterials is dramatically different from that in nanolayer structures, the spectra of both systems contain substantial contribution of sub-diffractional components.

\begin{figure}[t]
\centerline{\includegraphics[width=8cm]{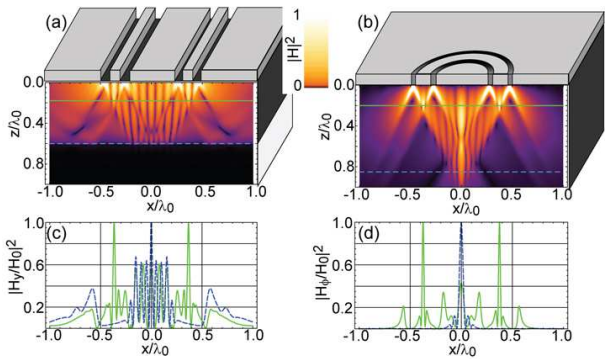}}
\caption{
(color online) (a) sub-diffractional light focusing at the interface between a nanolayer composite and air and in the bulk of the structure; (b) light focusing inside $2D$ hypergrating; profile of several sub-wavelength foci from (a) and (b) is shown in (c) and (d) respectively; the fields in (c,d) are normalized to the maximum field at the given distance from the grating; dashed and solid lines in (a,b) correspond the positions of field profiles shown in (c,d);   
geometry parameters: (a,c): $d_1=d_2=d_3=d_4=200 nm$, $x_1=-7\mu m; x_2=-4 \mu m, x_3=4 \mu m, x_4=7 \mu m$; (b,d): $d_1=d_2=200nm$, $r_1=4\mu m, r_2=7\mu m$
} 
\end{figure}

The light propagation behind an arbitrary 1D diffraction grating with slits of thickness $d_1, d_2, \ldots$ positioned at $x_1, x_2, \ldots$ is given by Eqs.(\ref{eqFld}\ldots\ref{eqTfun1}) with $S(k_x)=\sum_i S(k_x;x_i,d_i)$. Thus, each slit of the diffraction grating generates a set of two sub-wavelength beams diverging at the angle $\theta$. The interference of these beams can then be used to generate a sub-wavelength pattern in the bulk of metamaterial or on its opposite edge. 

The class of nanophotonic devices based on combinations of planar anisotropic (hyperbolic) metamaterials and diffraction gratings can be called {\it hyper-gratings}, reflecting some analogy between these systems, Fresnel optics and hyperlenses\cite{bornWolf,nfZonePlates,hyperlens} 

The developed framework is easily extendable for 2D gratings, with each point of the grating generating a cone of radiation with apex angle $2\theta$ [see Eq.\eqref{eqTheta}]. The superposition of these cones explains, in particular, the appearance of periodic sub-wavelength pattern in recent numerical simulations of imaging of $2D$ arrays of holes by multi-layered composites\cite{zhangLayers}. The limit $|\epsilon_z|\gg|\epsilon_{xy}|$ yields $\theta\rightarrow 0$ and corresponds to canalization regime observed in low-frequency nanowire structures\cite{belov}.

Hyper-gratings have potential to enable numerous exciting applications, including the communication between diffraction-limited optics and sub-wavelength areas, communication between several sub-wavelength objects inside the system, high-resolution lithography, and high-density sensing. 

In particular, hyper-gratings can be designed to realize the planar Fresnel-like lenses with subwavelength foci. Note that although the resolution of these lenses will still be limited by the spectral properties of incident radiation, hyper-gratings can provide unprecendented density of pixels at the focal plane. 

Several examples of generation of sub-wavelength field patterns inside (and at the back boundary) of the metamaterial are shown in Figs.2,3. Note that the thickness of the foci of the hyper-gratings ranges from $\lambda_0/50$ to $\lambda_0/20$ and is almost unaffected by the presence of material interfaces. 

We now turn to the analysis of limitations of the proposed technique. The main limitation of the resolution of anisotropy-based nanophotonics comes from material absorption. In the limit of low loss the high-$k_x$ behavior of the transfer function can be characterized by: 
\begin{eqnarray}
\label{eqTloss}
|T(k_x,z)|\simeq \exp\left[-\frac{z}{2}
\sqrt{-\frac{\epsilon_{xy}^\prime}{\epsilon_z^\prime}}
\left(\frac{\epsilon_{xy}^{\prime\prime}}{|\epsilon_{xy}^\prime|}+
\frac{\epsilon_{z}^{\prime\prime}}{|\epsilon_{z}^\prime|}\right)
k_x\right].
\end{eqnarray}
The evolution of the spectrum of the wavepackets inside anisotropic metamaterials calculated using numerical solutions of Maxwell's equations are shown in Fig.1. Note that the performance of realistic metamaterials greatly exceeds the performance of their isotropic counterparts. 

Assuming that the wavevector spectrum at the focal spot is dominated by the exponential decay given by Eq.\eqref{eqTloss}, and neglecting the specific dynamics of $|k_x|\lesssim\omega/c$ waves, we arrive to the following estimate of resolution of anisotropy-based hyper-gratings: 
\begin{eqnarray}
\label{eqRes}
\Delta\sim z
\sqrt{-\frac{\epsilon_{xy}^\prime}{\epsilon_z^\prime}}
\left(\frac{\epsilon_{xy}^{\prime\prime}}{|\epsilon_{xy}^\prime|}+
\frac{\epsilon_{z}^{\prime\prime}}{|\epsilon_{z}^\prime|}\right).
\end{eqnarray}
Our analysis indicates that under realistic conditions, Eq.\eqref{eqRes} tends to over-estimate the FWHM of the image. 

The second limitation of the proposed technique emanates from the appearance of non-local corrections in $k_x\gg \omega/c$ response of composite systems. Extensive previous research\cite{podolskiyWires,nonlocal} indicates that these corrections become important when the scale of field variation in the system becomes comparable with the size of its structural unit. Thus, the non-local corrections will limit the resolution of hyper-grating systems to the scale of the metamaterial component ($\sim 25\ldots 50 nm$ in realistic nanowire and nanolayer structures).

To conclude, we have proposed a new class of planar nanophotonic systems: hyper-gratings that combine the benefits of planar zone plates with far-field subwavelength resolution of the hyperlens. Sub-wavelength ``focal'' spots in the far-field of the hyper-grating have been demonstrated numerically and the analytical description of the underlying physic has been derived analytically. Examples of $1D$ and $2D$ amplitude gratings were presented. It is reasonable to assume that the results can be further optimized with phase gratings. The technique, illustrated here on examples of near-IR and mid-IR frequencies, is scalable from near-UV to mid-IR. Furthermore, the approach can be straightforwardly extended to enable the communication between sub-wavelength spots inside the bulk of metamaterials. Applications of the developed formalism lie in high-resolution sensing, lithography, and in on-chip communications.

This work has been supported by ONR (grant \#N00014-07-1-0457) and NSF (grant \#ECCS-0724763)



\begin{thebibliography}{}

\bibitem{shalaevNature}V.M. Shalaev Nat.Phot. {\bf 1}, 41 (2007)

\bibitem{pendryTransform} L.S. Dolin, Izv. VUZov, Radiofizika, {\bf 4}, 964 (1961); J. Pendry, D. Schurig, D. Smith, Science {\bf 312}, 1780 (2006); A. Kildishev, V.M. Shalaev, \ol {\bf 33}, 43 (2008)

\bibitem{hyperlens}Z. Jacob, L. Alekseyev, E. Narimanov, Opt.Exp., {\bf 14}, 8247 (2006); A. Salandrino, N. Engheta, Phys.Rev.B, {\bf 74}, 075103 (2006), Z. Liu, et.al., Science {\bf 315}, 1686 (2007); I.I. Smolyaninov, Y.J. Hung, C.C. Davis, Science {\bf 315}, 1699 (2007)

\bibitem{superlens} J.B.~Pendry, Phys. Rev. Lett. {\bf 85}, 3966 (2000); A. Grbic, G.Eleftheriades, \prl {\bf 92}. 117403 (2004); N.Fung, H.Lee, C.Sun, X.Zhang, Science {\bf 308}, 534 (2005); R.J.Blaikie, D.O.S.Melville, J.Opt.A. {\bf 7}, S176 (2005)

\bibitem{stockmanPRL} M.I. Stockman, Phys. Rev. Lett, {\bf 93}, 137404 (2004)

\bibitem{nfZonePlates}R.Merlin, Science {\bf 317} 927, (2007); A. Grbic, L. Jiang, R. Merlin, Science {\bf 320}, 511 (2008); L. Makley, A.M.H. Wong, Y.Wang, G.Eleftheriades, \prl {\bf 101}, 113901 (2008)

\bibitem{zhangLayers}Y. Xiong, Z.Liu, X.Zhang \apl {\bf 93}, 111116 (2008)

\bibitem{bornWolf} M. Born, E. Wolf, {\it Principles of optics}, Cambridge U. Press (Cambridge, UK 1999)

\bibitem{veselago} V.G. Veselago, Sov. Phys. Usp. {\bf 10}, 509 (1968); V.Veselago, E.E. Narimanov, Nat.Mat. {\bf 5}, 759 (2006)

\bibitem{podolskiyPRB} V.A.~Podolskiy and E.E.~Narimanov, Phys.~Rev.~B, {\bf 71}, 201101(R) (2005); A.A. Govyadinov, V.A. Podolskiy, \prb {\bf 73}, 155108 (2006)

\bibitem{indefinite}D. Schurig and D.R. Smith, New J. Phys. {\bf 7}, 162 (2005); G.X. Li, H.L. Tam, F.Y. Wang, and K.W. Cheah, \apj {\bf 102}, 116101 (2007) 

\bibitem{footnoteSiTheBest} The larger the dielectric constant of a material, the better the achievable resolution. Thus, since Si has one of the largest dielectric constants of all transparent natural materials, isotropic Si-based photonic system, shown in our simulation in a sense represent the best possible resolution 

\bibitem{natureSNOM}A.~Lewis, H.Taha, A.~Strinkovski, et.al. Nature Biotechnology, {\bf 21}, 1378 (2003)

\bibitem{podolskiyOL}  V.A. Podolskiy, E.E. Narimanov, \ol {\bf 30}, 75 (2005) 

\bibitem{hoffman} 11.	A. J. Hoffman, L. Alekseyev, S.S. Howard, K.J. Franz, D. Wasserman, V.A. Podolskiy, E.E. Narimanov, D.L. Sivco, C. Gmachl, Nat. Mat., {\bf 6}, 946 (2007)

\bibitem{podolskiyWires} J. Elser, R. Wangberg, V.A. Podolskiy, E.E. Narimanov \apl {\bf 89}, 261102 (2006) 

\bibitem{zhangWires}J.Yao, Z. Liu, Y.Liu, Y.Wang, C.Sun, G. Bartal, A. Stacy, X. Zhang, Science {\bf 321} 930 (2008)

\bibitem{zayatsWires}C. Reinhardt, S. Passinger, B.N. Chichkov, W. Dickson, G.A. Wutz, P.Evans, R.Pollard, A.V. Zayats \apl {\bf 89} 231117 (2006)

\bibitem{palik}E. Palik (ed.) {\it The handbook of optical constants of solids}, Academic Press (1997)


\bibitem{prime}in this work single ($^\prime$) and double ($^{\prime\prime}$) primes denote real and imaginary parts of complex numbers respectively

\bibitem{belov}P.A. Belov, C.R. Simovski, P.Ikonen, \prb {\bf 71}, 193105 (2005)

\bibitem{nonlocal}J. Elser, V.A. Podolskiy, I. Salakhutdinov, I. Avrutsky,\apl {\bf 90}, 191109 (2007); A.L. Pokrovsky,A.L.Efros, \prb {\bf 65}, 045110 (2002) 

\end{thebibliography}
\end{document}